\documentclass[13pt,a4paper]{article}
\usepackage[latin1]{inputenc}
\usepackage{amsmath}
\usepackage{amsfonts}
\usepackage{amssymb}
\usepackage{graphics}
\usepackage{graphicx}

\title{CP violation and mass hierarchy at medium baselines in the large $\theta_{13}$ era}

\begin{document}
\newcommand{\dcp}{$\delta_{CP}$~}
\newcommand{\tot}{$\theta_{13}$~}
\newcommand{\sot}{$\sin^2 2\theta_{13}$~}
\newcommand{\nue}{$\nu_e$~}
\newcommand{\nuebar}{$\bar{\nu}_e$~}
\newcommand{\nmcc}{$\nu_\mu^{CC}$~}
\newcommand{\numunue}{$\nu_\mu \rightarrow \nu_e$}
\newcommand{\numunutau}{$\nu_\mu \rightarrow \nu_\tau$}


\begin{center}
{\Large{\bf CP violation and mass hierarchy at medium baselines \\ in the large $\theta_{13}$ era} }\\
\vspace{.5cm}
S.~Dusini$^{\rm a}$, A.~Longhin$^{\rm b}$, M.~Mezzetto$^{\rm a}$,
L.~Patrizii$^{\rm c}$, M.~Sioli$^{\rm c,d}$, G.~Sirri$^{\rm c}$,  F.~Terranova$^{\rm e,f}$ 

\vspace*{1cm}
\noindent
$^{\rm a}$ Istituto Nazionale di Fisica Nucleare, Sez. di Padova, Padova, Italy \\
$^{\rm b}$ Laboratori Nazionali di Frascati dell'INFN, Frascati (Rome), Italy \\
$^{\rm c}$ Istituto Nazionale di Fisica Nucleare, Sez. di Bologna, Bologna, Italy \\
$^{\rm d}$ Dep. of Physics, Univ. di Bologna Bologna, Italy \\
$^{\rm e}$ Dep. of Physics, Univ. di Milano-Bicocca, Milano, Italy \\
$^{\rm f}$ Istituto Nazionale di Fisica Nucleare, Sez. di Milano-Bicocca, Milano, Italy \\
\end{center}

\abstract{The large value of \tot recently measured by rector and
  accelerator experiments opens unprecedented opportunities for
  precision oscillation physics. In this paper, we reconsider the
  physics reach of medium baseline superbeams. For \tot $\simeq
  9^\circ$ we show that facilities at medium baselines -- i.e. $L \simeq
  \mathcal{O} (1000 \mathrm{km}$) -- remain optimal for the study of CP
  violation in the leptonic sector, although their ultimate precision
  strongly depends on experimental systematics. This is demonstrated
  in particular for facilities of practical interest in Europe: a CERN
  to Gran Sasso and CERN to Phy\"asalmi $\nu_{\mu}$ beam based on the
  present SPS and on new high power 50~GeV proton driver. Due to the
  large value of \tot, spectral information can be employed at medium
  baselines to resolve the sign ambiguity and determine the neutrino
  mass hierarchy. However, longer baselines, where matter effects
  dominate the \numunue\ transition, can achieve much stronger
  sensitivity to $\mathrm{sign}(\Delta m^2)$ even at moderate
  exposures. \ }


\section{Introduction}

After an experimental search lasted more than a
decade~\cite{T2K,DoubleChooz,DayaBay,RENO}, we know quite precisely
the size of the mixing angle between the first and third neutrino
family ($\theta_{13}$)
\cite{An:2012eh,Ahn:2012nd,DoubleChooz_res,T2K_res}. Such value turned
out to be extremely large - in fact, very close to previous limits set
by CHOOZ~\cite{chooz} and Palo Verde~\cite{palo_verde}. A large value
of \tot not only constraints models for neutrino mass and
mixing~\cite{Altarelli:2004za} but, even more, opens up the
possibility to perform precision physics in the leptonic sector with
artificial neutrino beams in a way that resembles what was done for
quark mixing in the B-factory era~\cite{Mezzetto:2010zi}. In the past,
several experimental strategies have been
considered~\cite{Bandyopadhyay:2007kx,Battiston:2009ux}, corresponding
to possible values of \tot. The most extreme, addressing \sot as small
as $10^{-4}$, required novel acceleration techniques to achieve
unprecedented neutrino beam intensities and purity. The results from
T2K, Daya-Bay, RENO and Double-Chooz allow to reconsider these
scenarios, seeking for less challenging experimental setups to
establish CP violation in the leptonic sector and determine the
neutrino mass pattern (``mass hierarchy''). The experimental proposals
will take advantage of the large \numunue\ oscillation probability due
to $\theta_{13} = 8.9^\circ \pm 0.4^\circ$~\cite{Fogli:2012ua}, likely
relieving the constraints on the detector mass and accelerator
power. In this paper we address some of these issues, firstly on
general ground and, then, driven by practical considerations. On
general ground we discuss to what extent the constraints on beam
power, purity and systematic uncertainty can limit our capability to
establish CP violation in the leptonic sector using
Superbeams~\cite{Huber:2004ug,Barger:2007jq,Huber:2009cw,Coloma:2011pg,Coloma:2012wq,Coloma:2012ma}. We
consider the optimal baseline  to search for CP violation at
the atmospheric scale, i.e. a medium baseline ($L\simeq 1000$~km) with
neutrino energies of $E\simeq 1$~GeV and a far detector capable of
reconstructing quasi-elastic, deep-inelastic and resonance final
states (liquid argon TPC). It is well known that such baseline
represents a poor choice for the simultaneous determination of CP
violation (CPV) and mass hierarchy, which could be better achieved with
longer baselines and wide-band neutrino
beams~\cite{Barger:2006vy,Akiri:2011dv}.  Hence, in the context of
medium baselines we evaluate the deterioration of the sensitivity on
CPV due to the ignorance on the neutrino mass hierarchy. Similarly, we
compare the CP reach of the facility when the far detector is located
at a baseline optimal for the determination of mass hierarchy through
the exploitation of matter effects ($L \gg 1000$~km). The results of
such general study are relevant in setting experimental
issues and to establish a European strategy for precision
measurements in the large \tot era. 

First, we discuss to what extent the Gran Sasso underground
Laboratories could be considered as a viable candidate to host the
facility for CP violation. As an alternative
option~\cite{Baibussinov:2007ea}, we detail the physics reach of a new
shallow depth laboratory able to host significantly larger detector
masses than what can be accommodated in the present experimental halls
of LNGS. Finally, we compare the CP reach of these facilities with a
multipurpose detector~\cite{Autiero:2007zj,Rubbia:2009md} located at
longer baselines, i.e. from $\sim$1000 up to 2290 km
(CERN-to-Phy\"asalmi~\cite{Angus:2010sz,Rubbia:2010zz,Agarwalla:2011hh,talk_arubbia}),
in order to establish to what extent the choice of a baseline better
suited for the mass hierarchy can affect the CP reach. All comparisons
are done as a function of the intensity of the neutrino source,
detector mass and systematic uncertainties. When needed, beam
parameters are re-optimized ab initio (i.e. at the target-horn level)
to achieve best performance for a given configuration and allow a fair
comparison among various options.

\section{Facilities at medium baselines}

All current long-baseline experiments are in the medium 
baseline range ($\sim$ 1000~km) either because they are tuned to be at the peak of the
oscillation probability at the atmospheric scale (MINOS: 730 km, T2K:
293 km, NOVA: 810 km) or because they maximize the event rate for
\numunutau\ transitions (OPERA: 730 km). At these baselines, matter
density can be safely considered constant and \numunue\ transition
probabilities are commonly expressed by a perturbative expansion on
$\sin 2\theta_{13}$ and $\alpha \equiv \Delta m_{21}^2 / \Delta m_{31}^2 $. In
the following, $P(\nu_\mu \rightarrow \nu_e)$ is approximated
as~\cite{cervera_freund}:

\begin{eqnarray}
P(\nu_\mu \rightarrow \nu_e) & \simeq & \sin^2 2\theta_{13} \, \sin^2
\theta_{23} \frac{\sin^2[(1- \hat{A}){\Delta}]}{(1-\hat{A})^2}
\nonumber \\ & - & \alpha \sin 2\theta_{13} \, \xi \sin \delta
\sin({\Delta}) \frac{\sin(\hat{A}{\Delta})}{\hat{A}}
\frac{\sin[(1-\hat{A}){\Delta}]}{(1-\hat{A})} \nonumber \\ &+& \alpha
\sin 2\theta_{13} \, \xi \cos \delta \cos({\Delta})
\frac{\sin(\hat{A}{\Delta})}{\hat{A}} \frac{\sin[(1-\hat{A}){\Delta}]}
{(1-\hat{A})} \nonumber \\ &+& \alpha^2 \, \cos^2 \theta_{23} \sin^2
2\theta_{12} \frac{\sin^2(\hat{A}{\Delta})}{\hat{A}^2} \nonumber \\ &
\equiv & O_1 \ + \ O_2(\delta) \ + \ O_3(\delta) \ + \ O_4 \ \ .
\label{equ:probmatter}
\end{eqnarray}
In this formula $\Delta \equiv \Delta m_{31}^2 L/(4 E)$ and
the terms contributing to the Jarlskog invariant are split into the
small parameter $\sin 2\theta_{13}$, the ${\cal O}(1)$ term $\xi
\equiv \cos\theta_{13} \, \sin 2\theta_{12} \, \sin 2\theta_{23}$ and
the CP term $\sin \delta$; $\hat{A} \equiv 2 \sqrt{2} G_F n_e E/\Delta
m_{31}^2$ with $G_F$ the Fermi coupling constant and $n_e$ the
electron density in matter. Note that the sign of $\hat{A}$ depends on
the sign of $\Delta m_{31}^2$ which is positive (negative) for normal
(inverted) hierarchy of neutrino masses. 

The large size of \tot changes the phenomenology of the medium
baseline. In the past it was commonly believed $\sin 2\theta_{13}$ to
be at most of the size of $\alpha$, i.e. a large suppression of the
first term in Eq.\ref{equ:probmatter} was expected. We know now that
$\sin 2\theta_{13}$ overwhelms $\alpha$ by one order of magnitude ($\sim$0.3
versus $\sim$0.03), which makes the determination of the CP-blind term $O_1$
crucial for precision oscillation physics. Since $O_1$ is large and
depends on the mass hierarchy through $\hat{A}$, the ignorance on the
sign of $\Delta m_{31}^2$ plays a role even at medium baselines.

\begin{figure}[ht]
\centering
\includegraphics[scale=0.4,type=pdf,ext=.pdf,read=.pdf]{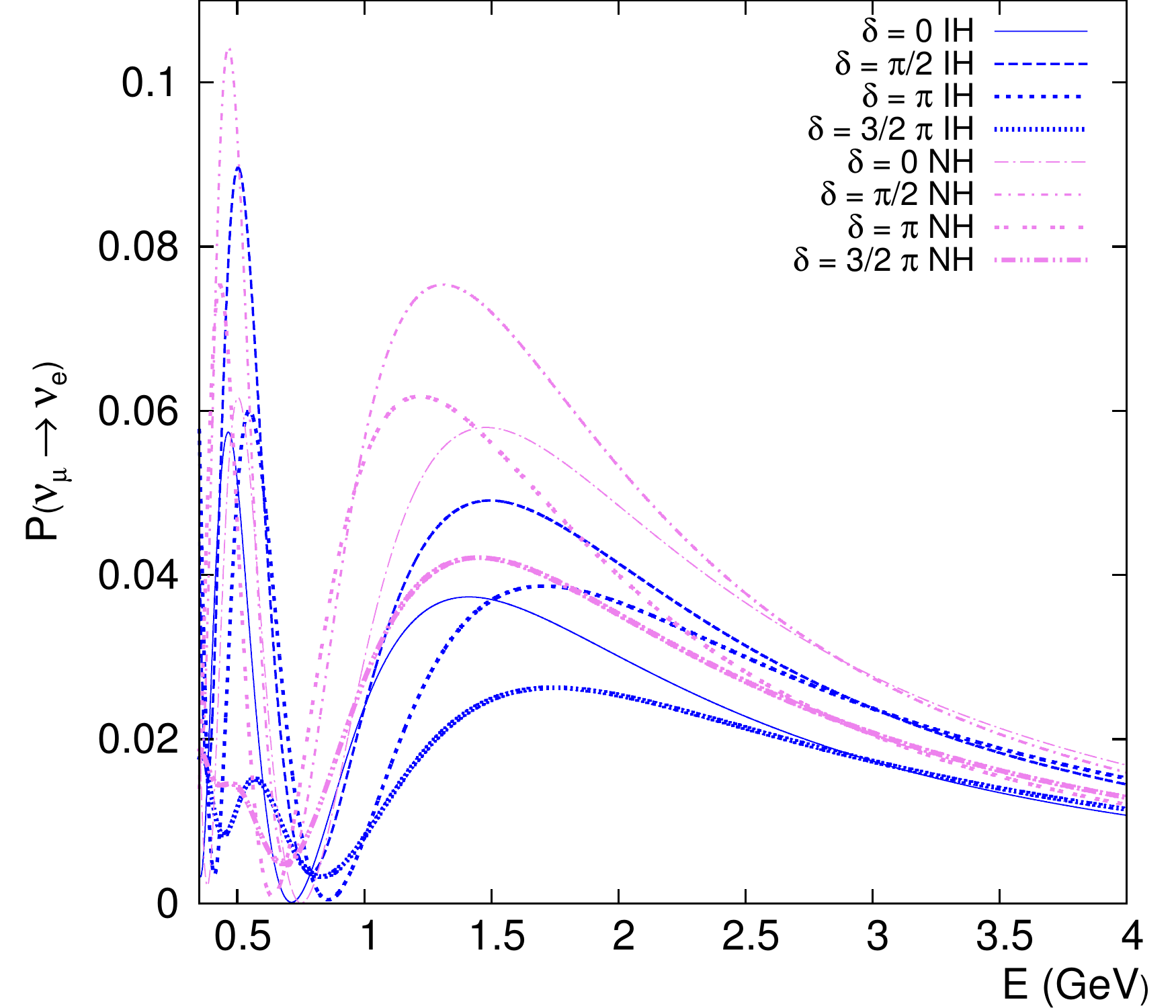}%
\includegraphics[scale=0.4,type=pdf,ext=.pdf,read=.pdf]{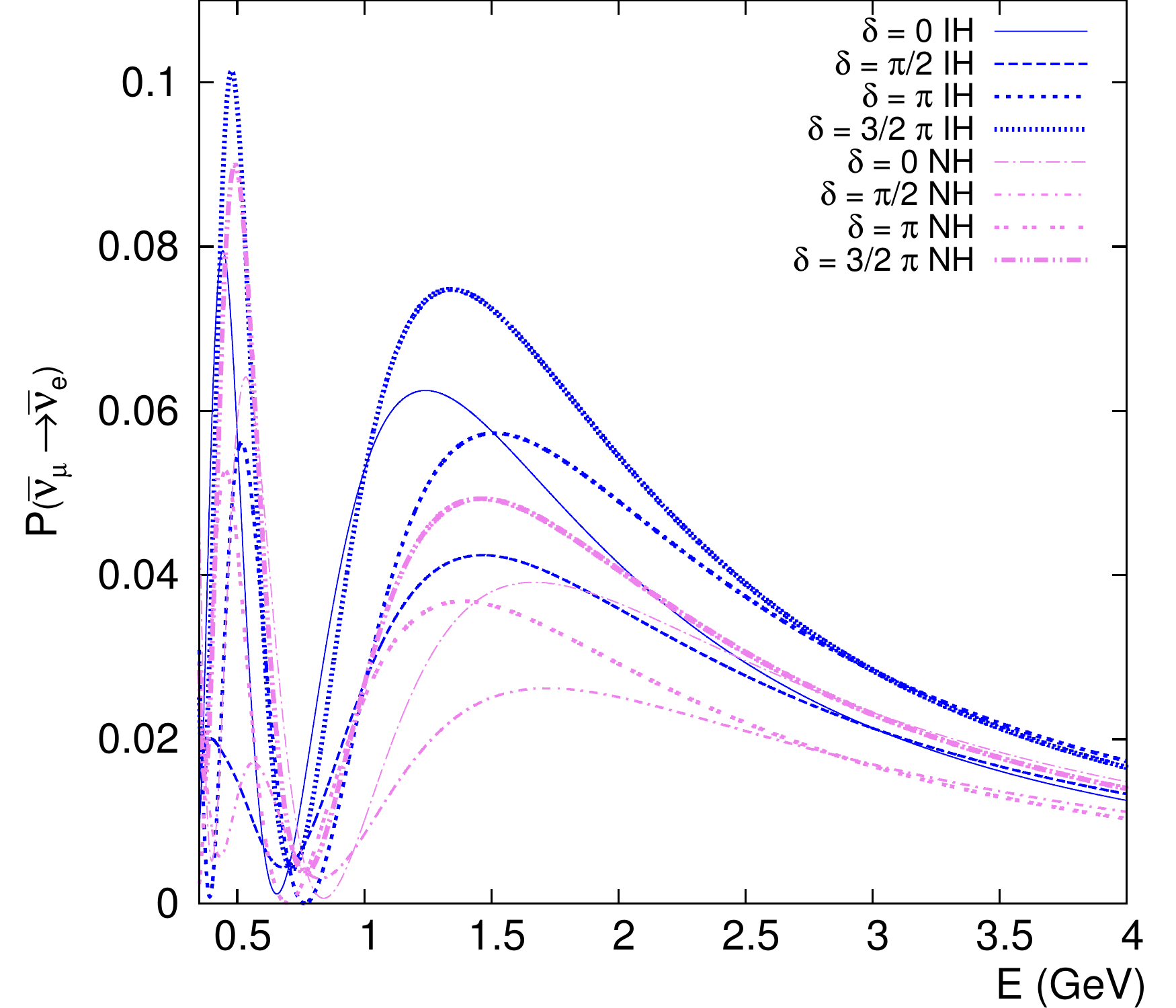} 
\caption{${\mathcal{P}}(\nu_\mu \to \nu_e)$ (left) and ${\mathcal{P}}(\bar{\nu}_\mu \to \bar{\nu}_e)$ (right) 
vs neutrino energy for different choices of the mass hierarchy and $\delta$ for a 730~km baseline and \sot $= 0.092$.}
\label{fig:fig00}
\end{figure}

For sake of illustration, Fig.~\ref{fig:fig00} shows the oscillation
probability ${\mathcal{P}}(\nu_\mu \to \nu_e)$ versus neutrino energy for
different choices of the mass hierarchy and the \dcp phase for a
baseline of 730~km and \sot $= 0.092$.  It is evident that effects
ruled by the mass hierarchy (matter effects) and the CP phase are of
comparable size. At first order they introduce a change in the
normalization so that the control of systematic errors becomes
crucial. The second oscillation maximum, that can be used to disentangle CP
and matter effects, appears at an energy of about 500 MeV. The full
exploitation of spectral information down to a few hundreds of MeV
points toward a detector with high granularity, energy resolution and
efficiency for low-energy electron neutrinos. That justifies the
consensus on the use of liquid argon (LAr) detectors as far detectors
for precision experiments at medium baselines (see
Sec.~\ref{sec:beamline_detector}).

In the following, three configurations will be detailed. All of them
exploit a neutrino beam from CERN to the Gran Sasso area and are based on a

\begin{enumerate}
\item on-axis detector inside the existing underground laboratory (up to 10 kt mass) (ONA) 
\item 7 km off-axis detector (up to 100 kt mass) located in a new shallow depth site (OA7)
\item as above but at 10 km off-axis (OA10)
\end{enumerate}

In particular, OA7 has been studied in the
past~\cite{Baibussinov:2007ea} for the measurement of $\theta_{13}$
although the mean neutrino energy is quite far from the oscillation
peak; on the other hand, the spectrum of OA10 better matches the
region of interest for CP violation at the price of reduced
statistics. The first option (ONA) is clearly of great practical
interest since it is completely based on existing underground
facilities. However, the limited size of the LNGS halls strongly
constraints the maximum detector weight~\cite{talk_votano}.  The
ICARUS T600 module~\cite{Amerio:2004ze,Rubbia:2011ft}, which is
currently in data taking in the Hall B of LNGS (110 m in length), has
a dimension of 3.9 $\times$ 4.3 $\times$ 19.6 m$^3$, corresponding to
a liquid argon mass of 0.735 kt (0.476 kt fiducial).  The ICARUS T1200
module \cite{T1200} proposed in 2001 for the CNGS was based on a tank
of 10.3 $\times$ 10.3 $\times$ 21 m$^3$, totaling 1.47 kt (0.952 kt
fiducial).  Assuming an array of four T1200 modules located in Hall B
would allow the installation of a 4 kt fiducial mass detector at Gran
Sasso. In addition, a novel design recently considered by the ICARUS
Collaboration can increase the total available mass in Hall B up to
7.5 kt \cite{RubbiaNUTURN}. In conclusion, even assuming the
availability of extra space at LNGS, the Gran Sasso laboratories could
not host a detector with mass larger than 10 kton. In this study 10
kton is considered the largest explotable mass for the ONA option.

Similarly, two classes of options have been considered for the neutrino
source (see Sec.~\ref{sec:beamline_detector}). The former, which is
based on the existing CERN-SPS, employs at most facilities that are already
available; the second one, which considers a new CERN-based proton driver
at 50~GeV, allows for higher flexibility in the choice of the beam
configuration and better performance on a longer timescale.

\section{Beamline and detector simulation}
\label{sec:beamline_detector}

\begin{figure}
\centering
\includegraphics[scale=0.35,type=pdf,ext=.pdf,read=.pdf]{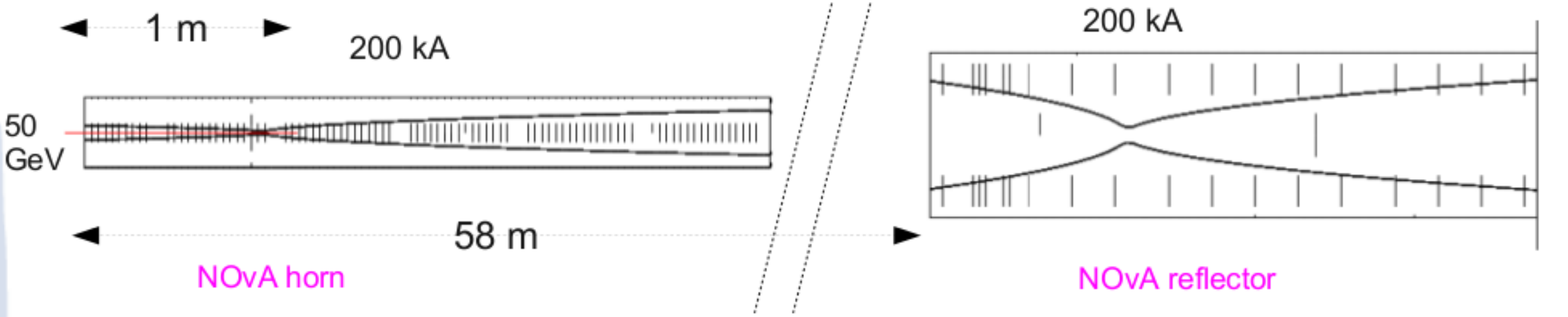}\\
{~~~~~~~~~~~~~~~~~~~~~}\\
{~~~~~~~~~~~~~~~~~~~~~}\\
\includegraphics[scale=0.35,type=pdf,ext=.pdf,read=.pdf]{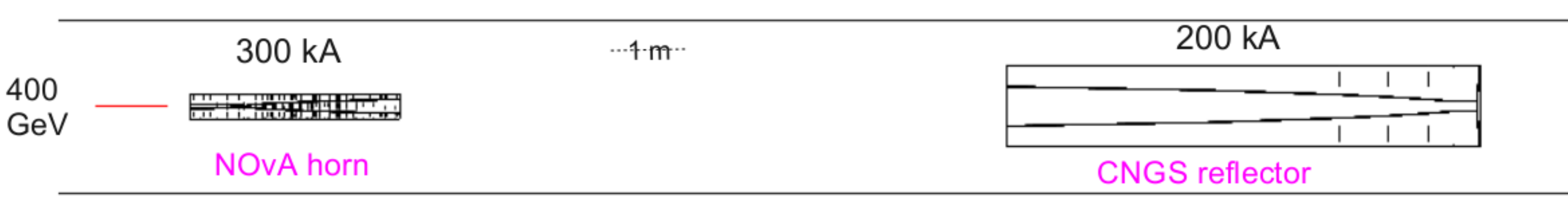} 
\caption{Focusing optics for the 50 (top) and 400 GeV (bottom) proton energy beamline.}
\label{fig:optics}
\end{figure}

The beam parameter optimization for a novel 50~GeV proton driver is
based on the results obtained in \cite{flussiLAGUNA} in the framework
of the LAGUNA design study. The most relevant parameters are the
distances among the target, the upstream focusing horn and the
downstream focusing horn (``reflector''), the target length and the
length and width of the decay tunnel. Following the approach of
Ref.~\cite{flussiLAGUNA}, the parameters, together with the geometry
of the horn and reflectors were optimized using the sensitivity to
\tot as a figure of merit, instead of resorting to intermediate
observables as e.g. the number of unoscillated neutrinos. This
optimization is also appropriate for the CP reach, especially if the
detector technology allows the exploitation of spectral
information. The optimal configuration for a $\sim 700$~km baseline
(CERN-to-Gran Sasso) is shown in the top drawing of Fig.~\ref{fig:optics}. 
The decay tunnel is 90~m long and the optimal radius is 2.2~m. The target
consists of a 1~m long graphite rod. Primary interactions in the
target were simulated with the GEANT4~\cite{geant4} QGSP hadronic
package and secondaries were traced down to the decay tunnel. The \tot
sensitivity was computed using the GLoBES 3.1.11
package~\cite{Huber:2004ka,Huber:2007ji} and assuming 100 kton LAr far
detector (see below).

\begin{figure}
\centering
\includegraphics[scale=0.5,type=pdf,ext=.pdf,read=.pdf]{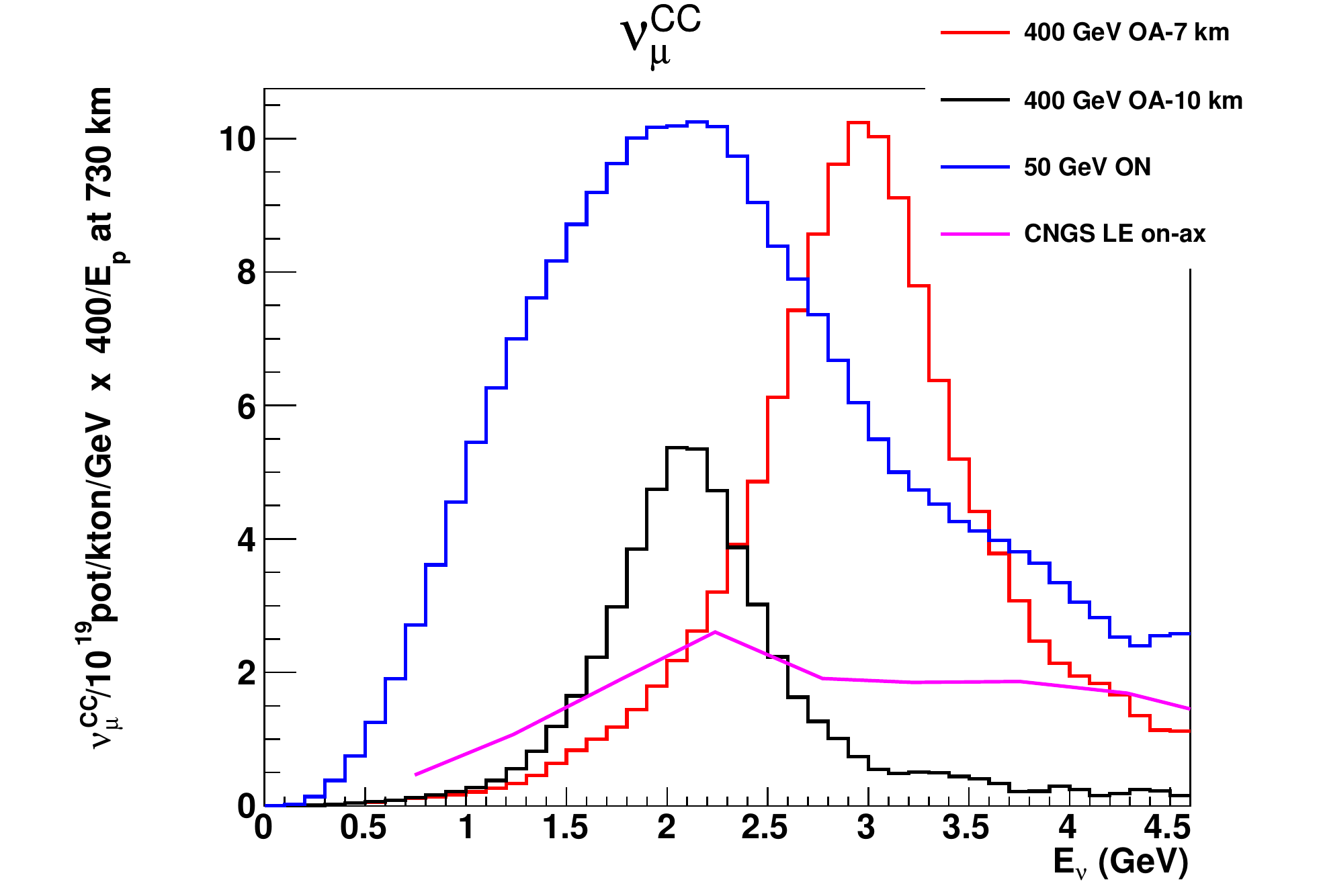} 
\caption{\nmcc event rates for the two off-axis (OA) options
considered in the text (``400 GeV OA 7-km'' and ``400 GeV OA 10-km'') and for
a dedicated on-axis 50 GeV proton driver (``50 GeV ON''). ``CNGS
LE on-ax'' refers to the low-energy (10 GeV optics)
re-optimization of CNGS considered in \cite{Meregaglia:2006du}.}
\label{fig:nuflux_onaxis_rate}
\end{figure}

The neutrino beam from the existing SPS required a dedicated
optimization since the current CNGS configuration (on-axis, $\langle
E_\nu \rangle \sim 17$~GeV) is not appropriate for CP violation
studies. An approach similar to the optimization of the 50~GeV driver
was followed. A pre-scan of the beam parameters was carried out using
a fast simulation based on the BMPT
parameterization~\cite{Bonesini:2001iz}. This approach is particularly
rewarding since the BMPT formulas are mostly based on data collected
with 400 GeV/c and 450 GeV/c protons at CERN-SPS. However, fluxes in
the proximity of the optimal parameters were fully simulated with
GEANT4. The optimised configuration is shown in the bottom drawing of
Fig.~\ref{fig:optics}.  The tunnel geometry is the one of CNGS with a
length of 1000 m and a radius of 1.225 m.  A 1 m long graphite target
was used.  The corresponding off-axis event rates in the optimal
configuration are shown in Fig.~\ref{fig:nuflux_onaxis_rate} together
with the rates for the ONA configuration (``50 GeV ON''). All rates are normalized
to the proton energy.

The advantage of a dedicated proton driver at 50~GeV is quite evident
since the region where most of the CP related information reside is
located below 2.5~GeV. Moreover, at these baselines the second
oscillation maximum is inaccessible for any configuration based on a
400~GeV accelerator. Similar considerations hold for a low energy (10
GeV optics) re-optimization of CNGS (``CNGS LE on-ax'' in
Fig.~\ref{fig:nuflux_onaxis_rate}), as the one considered in
Ref.~\cite{Meregaglia:2006du}.

\begin{figure}
\centering
\includegraphics[scale=0.33,type=pdf,ext=.pdf,read=.pdf]{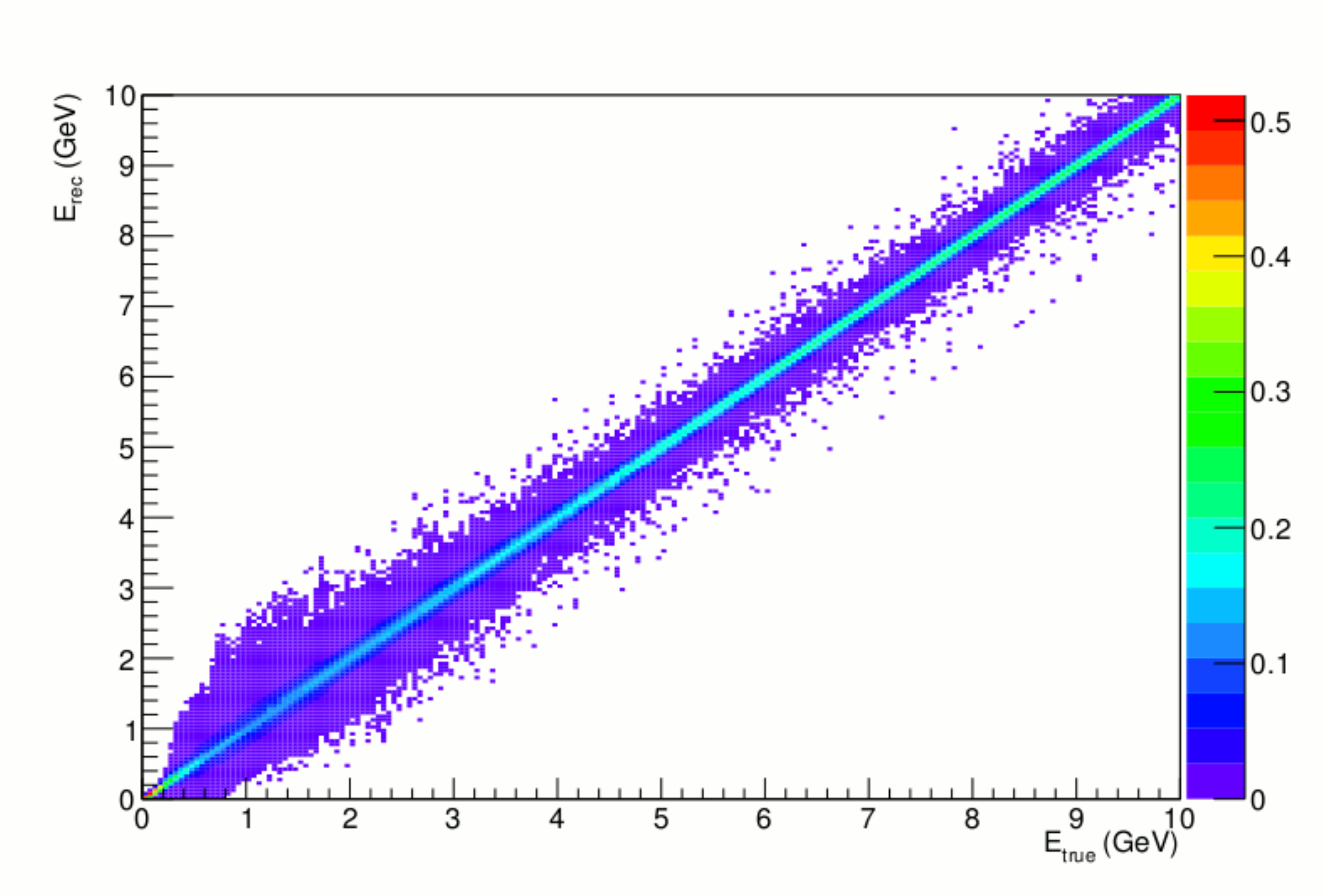}%
\includegraphics[scale=0.25,type=pdf,ext=.pdf,read=.pdf]{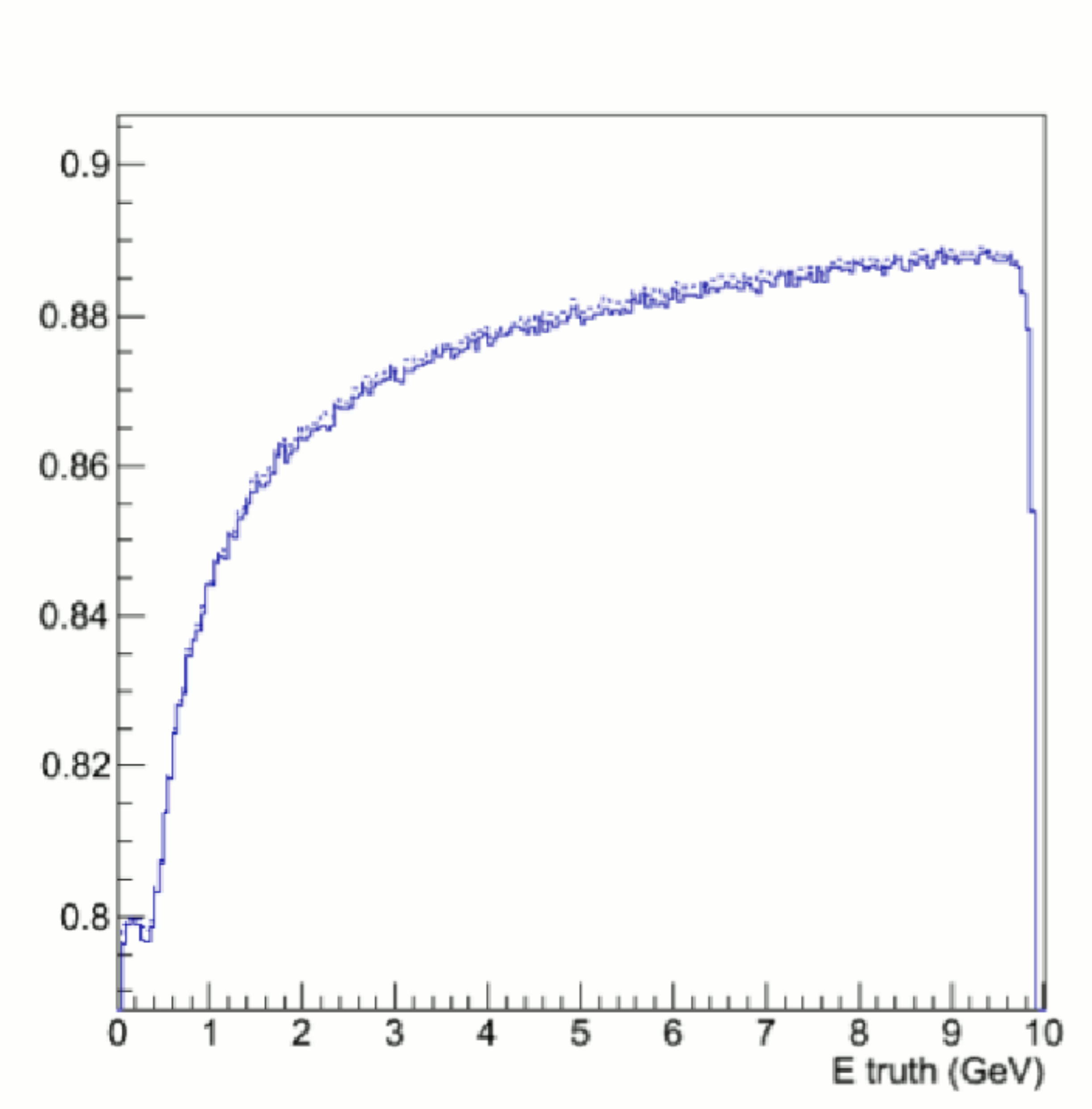} 
\caption{Smearing matrix (left) and efficiency (right) for \nue CC events.}
\label{fig:SMEARING}
\end{figure}

Liquid Argon detectors are the most promising technology to address
precision oscillation physics. The main advantages of a LAr TPC in
$\nu_e$ appearance are very high efficiency both for quasi-elastic
(80\%) and deep-inelastic (90\%) interactions combined with a superior
NC rejection power. In the following, we considered a contamination of
NC due to $\pi \rightarrow e$ misidentification not exceeding 0.1\% of
the $\nu_\mu$ CC rate. In the occurrence of QE interaction, the
neutrino energy can be fully reconstructed by the lepton energy and
direction, since the direction of the incoming neutrino is known in
advance. LAr
detectors~\cite{Amerio:2004ze,Rubbia:2011ft,Ankowski:2008aa} are
capable to reconstruct $E_\nu$ with a resolution mostly dominated by
the electron energy resolution: $\sigma_{E_\nu}/E_{\nu} \simeq 0.05/
\sqrt{E_\nu}$, $E_\nu$ being expressed in GeV. On the other hand,
energy resolution for DIS-$\nu_e$ interaction is driven by the
resolution on the hadronic system. In LAr, the latter amounts to
$\sigma_{E_h}/E_{h} \simeq 0.2/ \sqrt{E_h \mathrm{(GeV)}}$. In this
study, the energy resolution and efficiency for $\nu_e$ and
$\bar{\nu_e}$ were implemented smearing the final state momenta of the
electrons and hadrons. Interactions were simulated using the GENIE
Monte Carlo generator~\cite{Andreopoulos:2009rq} and the corresponding
migration matrices were implemented in the detector description of
GLoBES.  The smearing matrices were calculated for \nue and \nuebar
separately. They are shown for $\nu_e$ events in
Fig.~\ref{fig:SMEARING} (left). In the simulation, a 50 MeV bin width
was used.  Similarly, the efficiency as a function of energy is shown
in Fig.~\ref{fig:SMEARING} (right).

\section{Results}

\begin{figure}
\centering
\includegraphics[scale=0.4,type=pdf,ext=.pdf,read=.pdf]{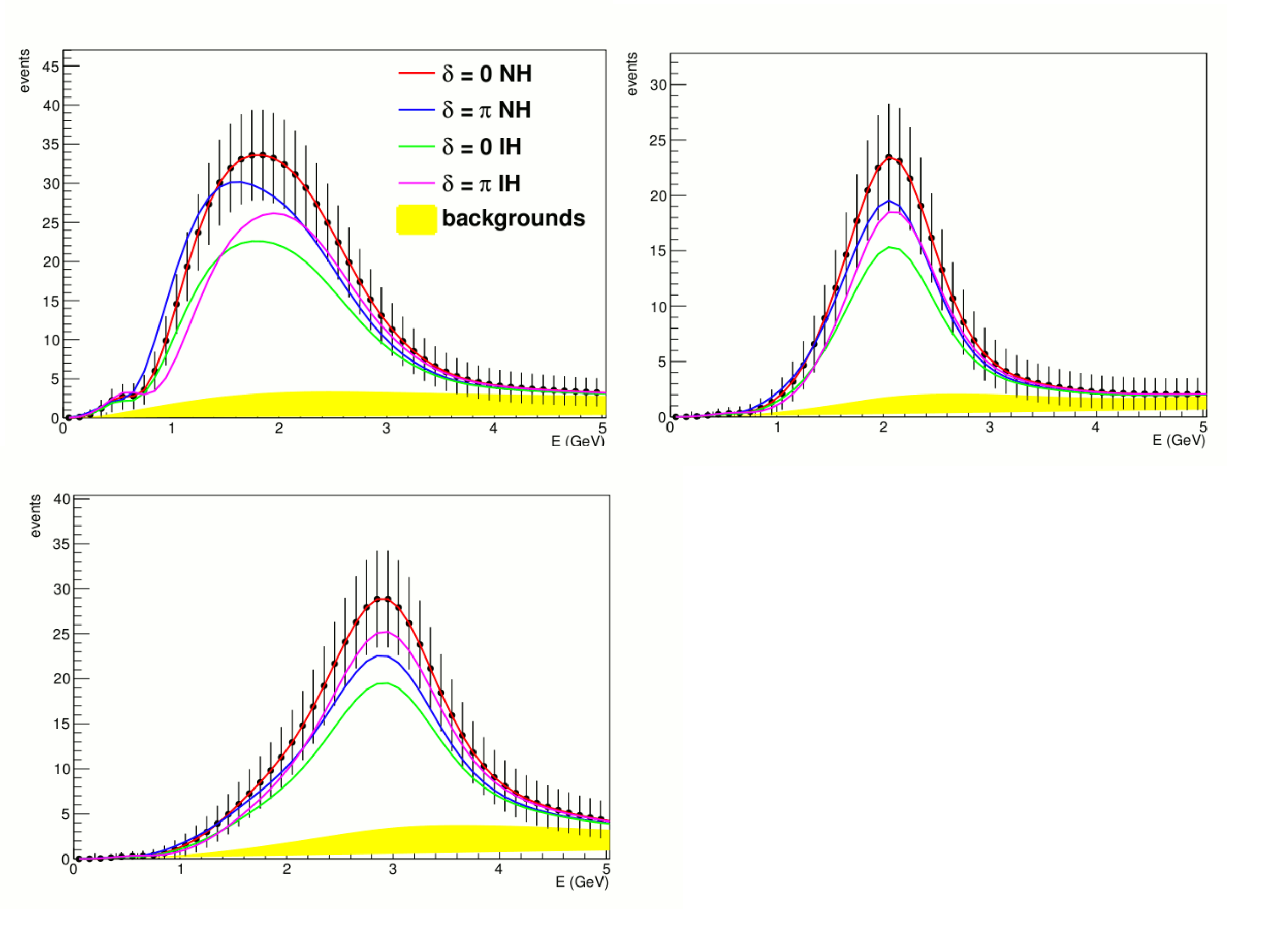}
\caption{\nue appearance signal  
  at 730 km for 5 years of data taking, $\sin^2 2\theta_{13}=0.092$,
  and normal hierarchy.  Upper Left: ONA option, 10 kt $\otimes$ 3
  $\cdot$ 10$^{21}$ pot/y (2.4 MW). Upper Right: OA10 option, 20 kt
  $\otimes$ 1.2 $\cdot$ 10$^{20}$ pot/y (0.77 MW). Lower Left: OA7 option ,
  20 kt $\otimes$ 1.2 $\cdot$ 10$^{20}$ pot/y (0.77 MW).  }
\label{fig:nuespectra}
\end{figure}

As mentioned above, three possible options were considered in this
study: an on-axis detector with limited fiducial mass (ONA), a 7~km
(OA7) and 10~km (OA10) off-axis setup leveraging the existing 400 GeV
accelerator. In Fig. \ref{fig:nuespectra} the $\nu_e^{CC}$ appearance
spectra convoluted with detector effects for the ONA, OA7 and OA10
options are shown. For the on-axis option, a 50 GeV proton driver with
a 10 kt detector was assumed. The power of the proton driver
corresponds to 3 $\cdot$ 10$^{21}$ pot/y (2.4 MW).  For the off-axis
options, a 20~kton detector and 1.2 $\cdot$ 10$^{20}$ pot/y (0.77 MW)
was considered. Clearly, for the 400~GeV proton driver option the power of the driver is
constrained by the limitations of the SPS-based neutrino beam, while
for the 50~GeV facility, a dedicated multi-MW machine can be envisioned.

The performance of the different options were compared defining as
figure of merit the CP violation discovery potential at
$3\sigma$. Following~\cite{Bandyopadhyay:2007kx,Huber:2002mx}, we
define the ``CP coverage'' at $3\sigma$ as the fraction of possible
(true) values of the CP phase $\delta$ where the CP conserving
hypothesis ($\delta=0,\pi$) has a p-value smaller than 0.0027. It
corresponds to $\chi_{min}>9$ for $\delta=0,\pi$ (1 d.o.f.). In each
case, we considered separately the possibility that the mass hierarchy
is known by the time the facility starts data taking. If the mass
hierarchy is not known, the p-value of the null hypothesis is computed
as the $\chi_{min}$ of the following combinations: ($\delta=0, \Delta
m^2_{32}>0$),($\delta=\pi, \Delta m^2_{32}>0$), ($\delta=0, \Delta
m^2_{32}<0$), ($\delta=\pi, \Delta m^2_{32}<0$).

\begin{figure}
\centering
\includegraphics[scale=0.5,type=pdf,ext=.pdf,read=.pdf]{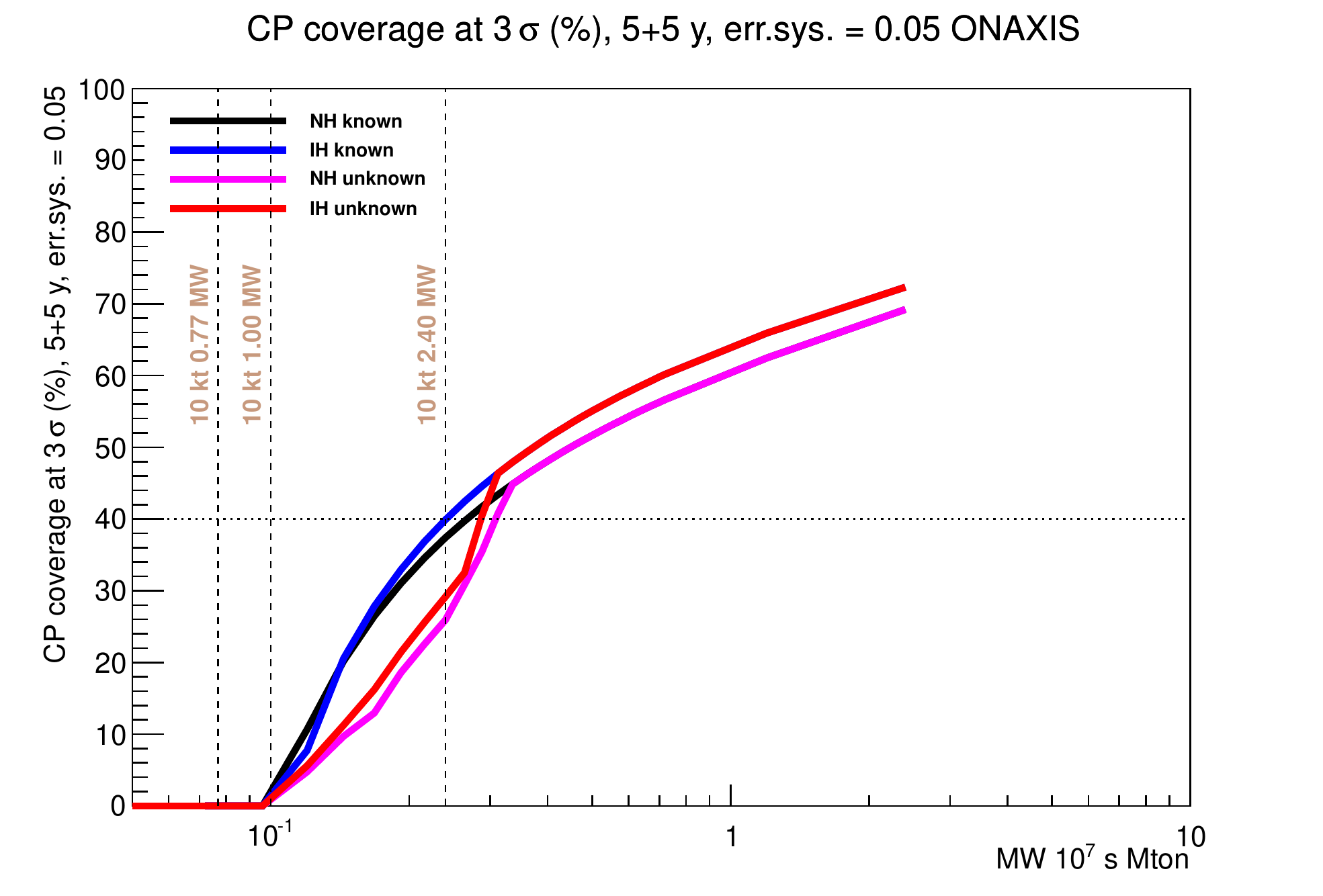}
\caption{CP coverage at $3\sigma$ for 5 years of $\nu$ and 5 years of
  $\bar{\nu}$ running with the on-axis configuration, a 50 GeV proton
  driver and a systematic error on flux normalization of 5\%. We
  consider normal and inverted hierarchy assuming this information to
  be either available or not (color codes).}
\label{fig:result1}
\end{figure}

\begin{figure}
\centering
\includegraphics[scale=0.5,type=pdf,ext=.pdf,read=.pdf]{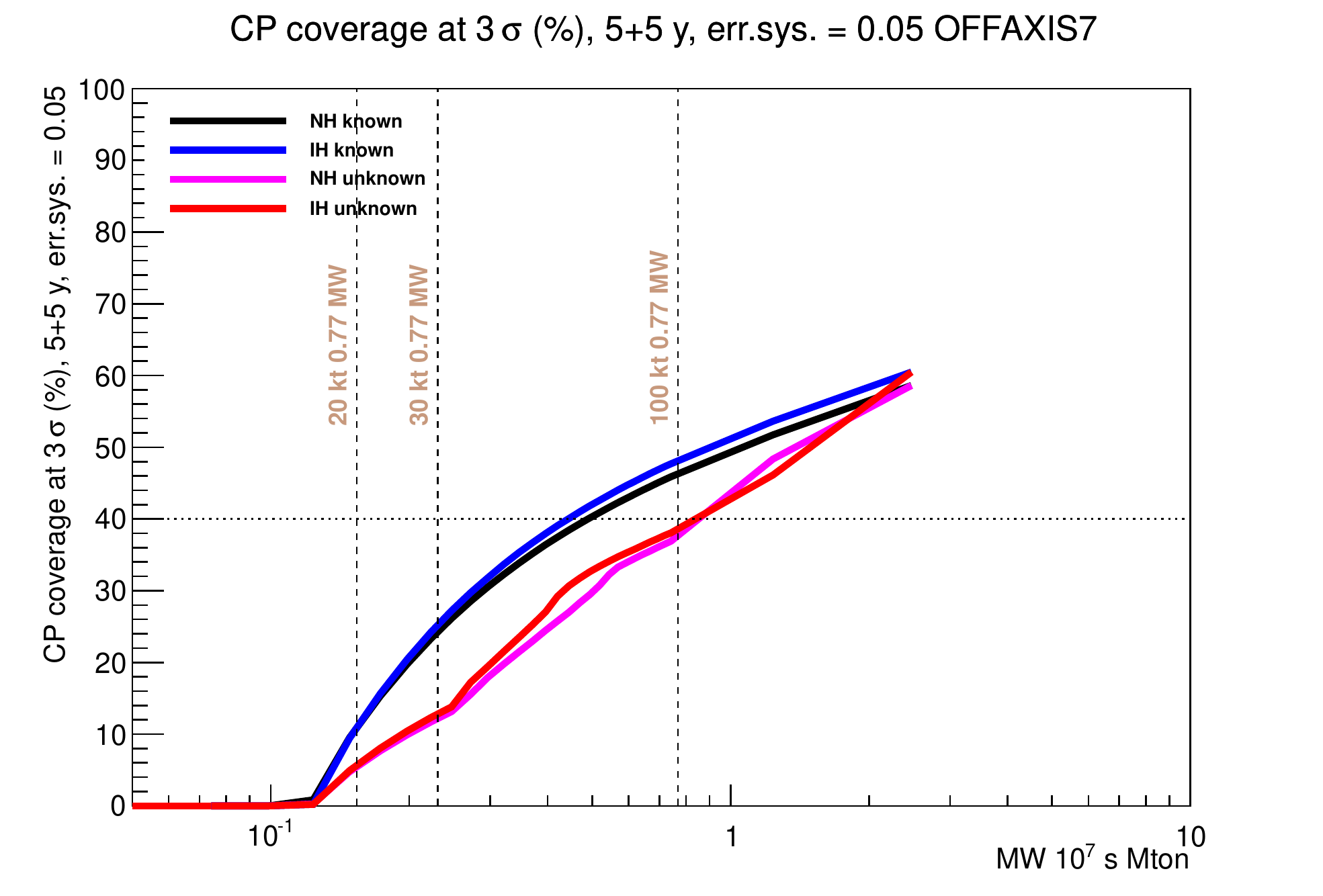}
\caption{CP coverage at $3\sigma$ for 5 years of $\nu$ and 5 years of
  $\bar{\nu}$ running with the off-axis 7 km configuration, a 400 GeV
  proton driver and a systematic error on flux normalization of
  5\%. We consider normal and inverted hierarchy assuming this
  information to be either available or not (color codes).}
\label{fig:result2}
\end{figure}

\begin{figure}
\centering
\includegraphics[scale=0.5,type=pdf,ext=.pdf,read=.pdf]{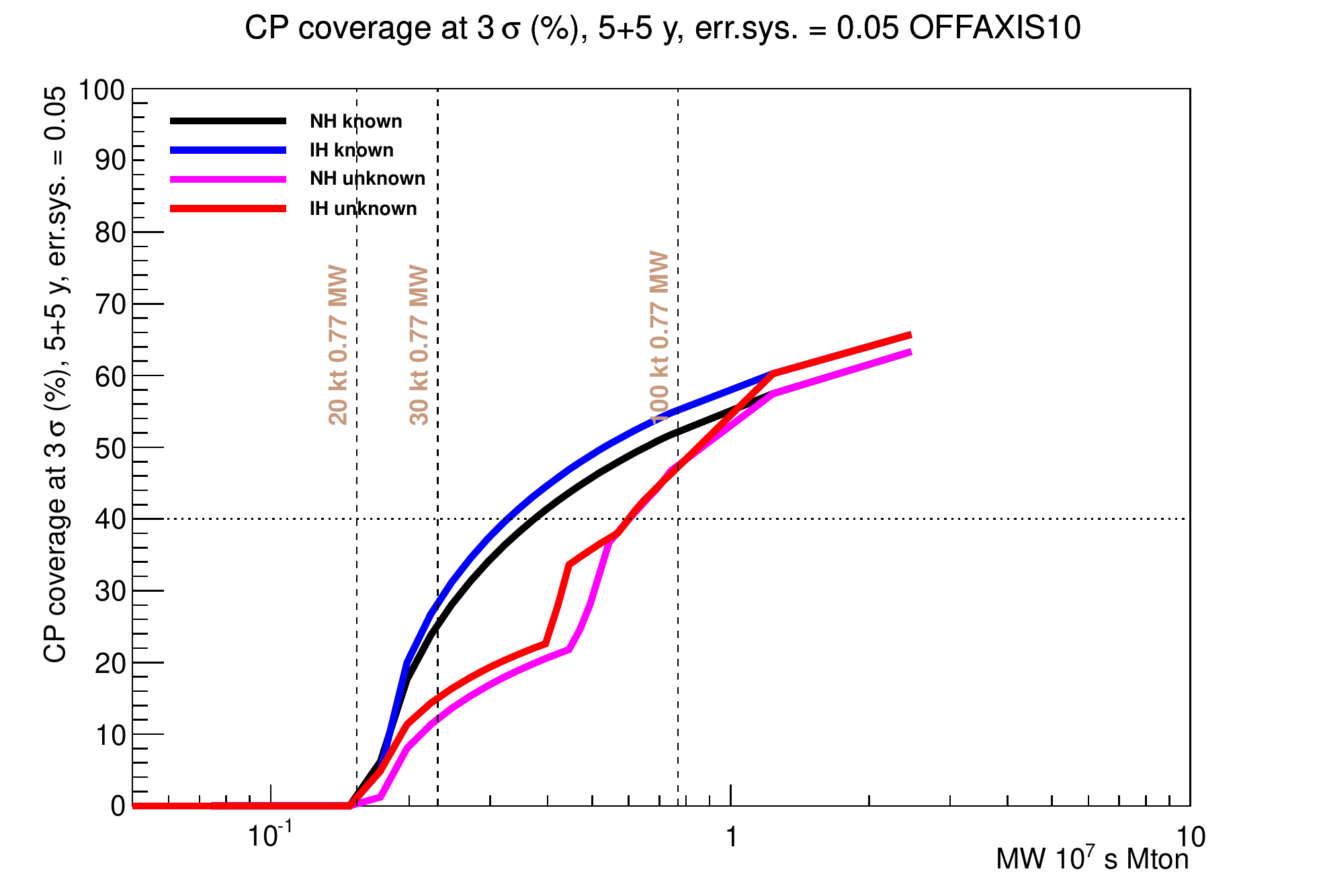}
\caption{CP coverage at $3\sigma$ for 5 years of $\nu$ and 5 years of $\bar{\nu}$ running
with the off-axis 10 km configuration, a 400 GeV proton driver and a systematic error on flux 
normalization of 5\%. We consider normal and inverted hierarchy assuming this information to be either available or not
(color codes).}
\label{fig:result3}
\end{figure}

The $3\sigma$ CP coverage in the on-axis facility based on the 50~GeV
proton driver is shown in Fig.~\ref{fig:result1} for 5 years of
running with $\nu$ and 5 years with $\bar{\nu}$. The corresponding
results are shown in Figs.~\ref{fig:result2} and ~\ref{fig:result3}
for the 7 and 10~km off-axis configurations based on the 400 GeV
proton driver.  In each plot we consider normal and inverted
hierarchy, assuming this information to be available or not (color
codes) at time of running. The assumed systematic error on flux
normalization amounts to 5~\%.

\begin{figure}
\centering
\includegraphics[scale=0.5,type=pdf,ext=.pdf,read=.pdf]{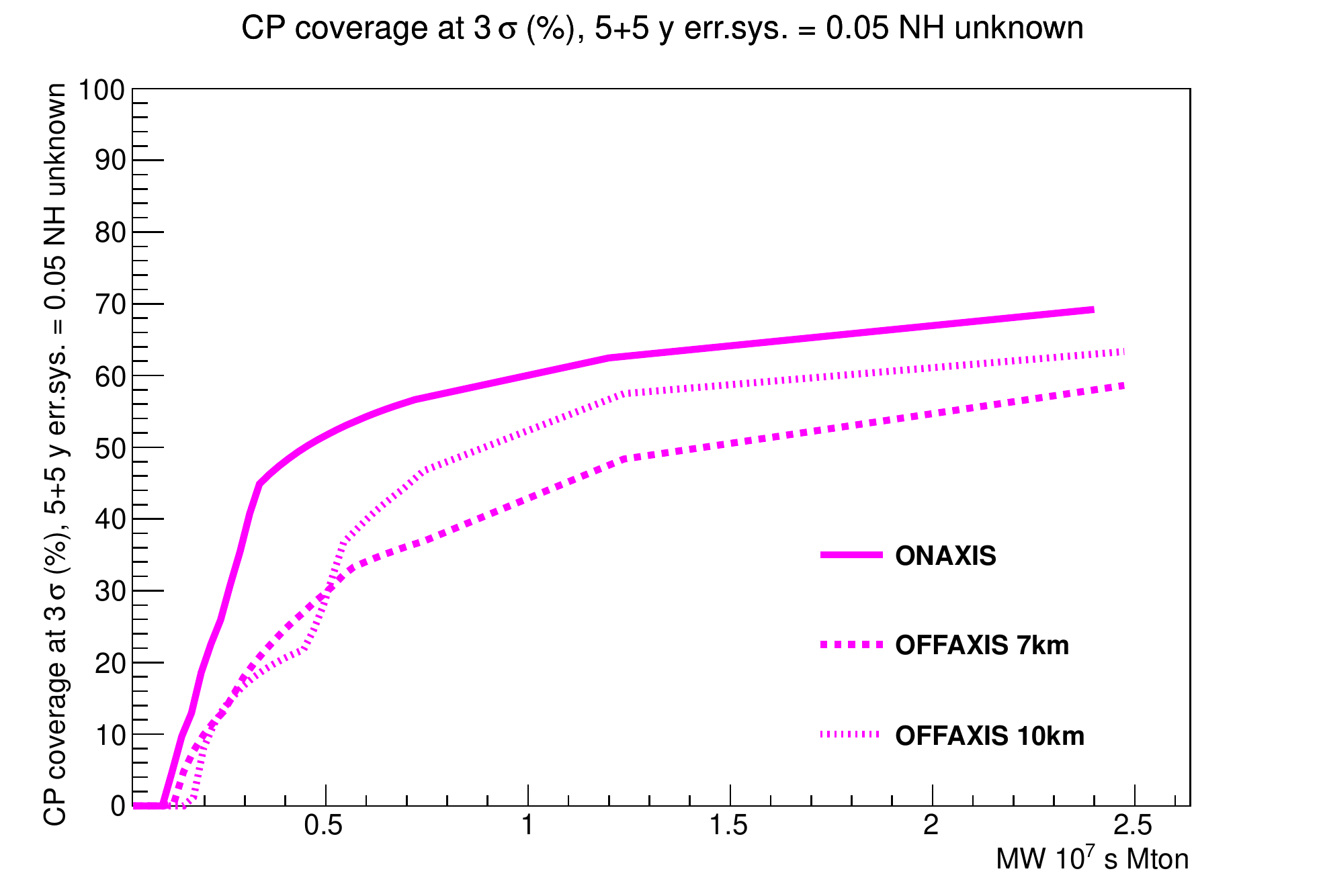}
\caption{
Comparison of the CP coverage at $3\sigma$ for 5 years of $\nu$ and 5 years of $\bar{\nu}$ running
with the on-axis 50 GeV configuration, off-axis 7 and 10 km configuration and 400 GeV proton driver, 
and a systematic error on flux  normalization of 5~\%. Here we only consider normal hierarchy, which
is assumed to be unknown.}
\label{fig:result4}
\end{figure}

A comparison of the three options in the case of normal hierarchy
(assumed to be unknown) is plotted in Fig. \ref{fig:result4}.

Some interesting considerations hold either when the results are
considered on general ground for any medium baseline facility or when
specific CERN-based facilities are envisioned. Firstly, it is clear
that the limitation due to the sign
degeneracy~\cite{Minakata:2001qm,Barger:2001yr}, i.e. the
deterioration of the CPV coverage due to the ignorance of the mass
hierarchy plays a major role only for limited exposures or detector
mass and it is not an intrinsic limitation of medium baselines for
such high values of $\theta_{13}$.  Large exposures, however, are
mandatory to exploit spectral information and thus to remove the sign
ambiguity at $L\simeq 730$~km.

\noindent
The constraints on the minimum exposure is particularly severe when we
consider realistic CERN-based facilities. In the vertical lines of
Figs.~\ref{fig:result1} we indicated the CP coverages of a facility
whose far detector is hosted in the underground halls of LNGS (maximum
LAr mass: 10 kton). It is apparent that LNGS is not appropriate as a
far detector site unless a dedicated proton driver well exceeding 2~MW
power can be built at CERN. Similar considerations hold for the
off-axis facilities that leverage the infrastructure on which the CNGS
is built (Fig.\ref{fig:result2},\ref{fig:result3}).  In this case, a
new shallow laboratory can overcome the limitation on the detector
mass but the low power of the driver constraints the CP coverage well
below 50\%. Again, the limited statistics reduce the resolution power
of spectral information to lift the sign ambiguity, so that prior
knowledge of the mass hierarchy increases the CP coverage up to a
factor 2.

\begin{figure}
\centering
\includegraphics[scale=0.39,type=pdf,ext=.pdf,read=.pdf]{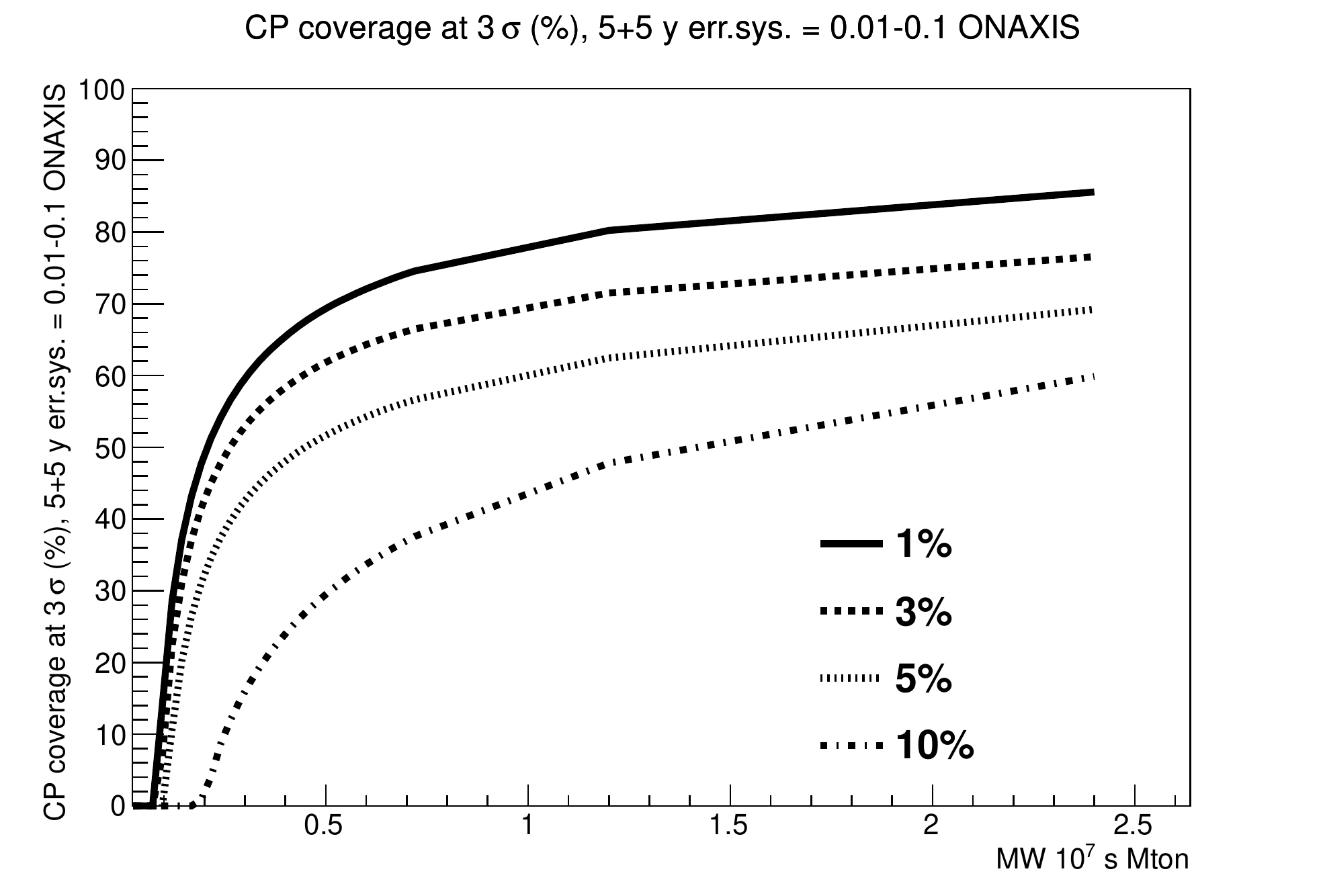}%
\includegraphics[scale=0.39,type=pdf,ext=.pdf,read=.pdf]{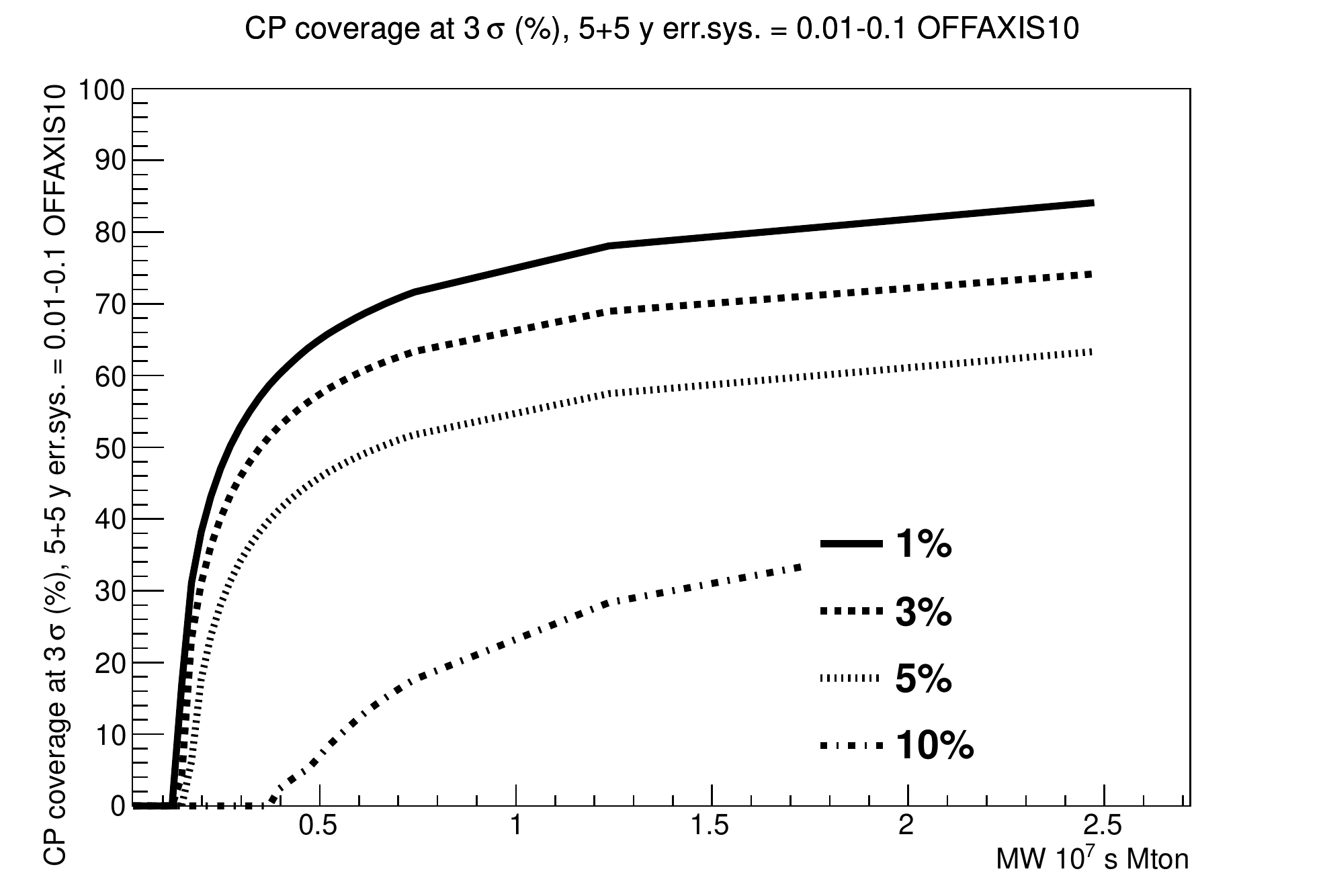}
\caption{Effect of systematics on the CP coverage for the on axis
  (left) and off-axis 10~km (right) configurations.}
\label{fig:sys}
\end{figure}

At large exposures the coverage quickly saturates at $\sim$70\% (see
Fig.\ref{fig:result4}). This is an intrinsic feature of Superbeams for
large $\theta_{13}$ where the CP reach is dominated by the systematics
on the flux normalization. That is shown in Fig.~\ref{fig:sys}, where
systematics are varied from 1\% to 10\%. In this range, the coverage
at saturation decreases from 86\% to 60\%. This result confirms the
key role played by the near detector and by the knowledge of the cross
sections and detector efficiency when performing precision physics
with Superbeams~\cite{Huber:2007em}. In fact, a control of the
systematics at the few-percent level remains very challenging, being still
at the level of 10.3\% for T2K. Clearly, this issue makes the
construction of a near detector mandatory for CPV searches. This is
particularly relevant for CERN-based facilities, where the
installation of a near detector poses additional difficulties from the
point of view of engineering and civil infrastructure.

\begin{figure}
\centering
\includegraphics[scale=0.55,type=pdf,ext=.pdf,read=.pdf]{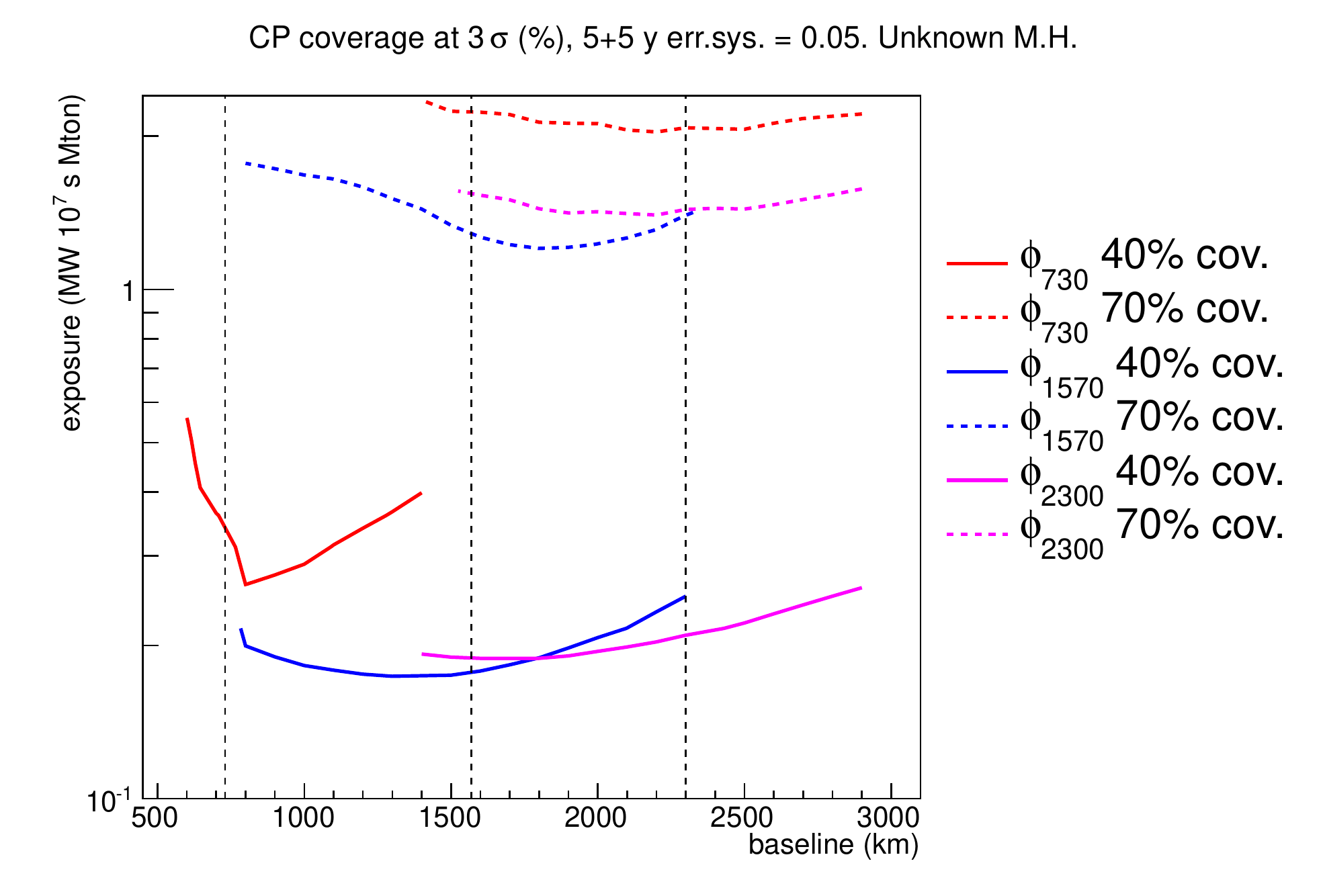}%
\caption{Exposure needed to reach 40\% (solid line) and 70\% (dashed
  line) CP coverage as a function of the baseline.  The red, blue and
  purple contours are drawn considering the optimal beamline (see
  Sec.~\ref{sec:beamline_detector}) for a 730, 1570 and 2300~km
  baseline, respectively.  }
\label{fig:compare_baseline}
\end{figure}

\begin{figure}
\centering
\includegraphics[scale=0.7,type=pdf,ext=.pdf,read=.pdf]{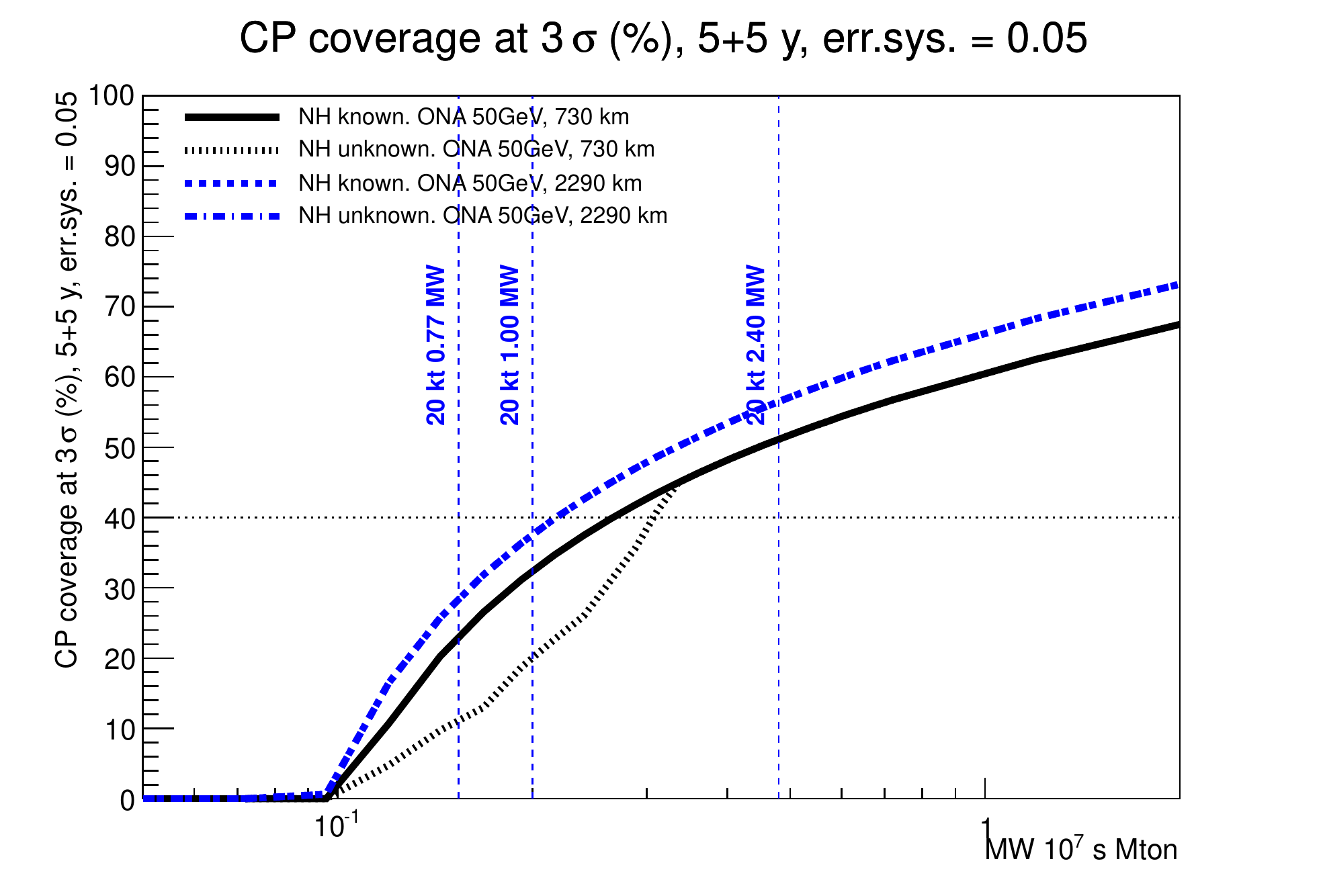} 
\caption{CP coverage for the facilities based on the 50 GeV proton driver (see text for details).}
\label{fig:LNGS-Finland_0}
\end{figure}

\begin{figure}
\centering
\includegraphics[scale=0.7,type=pdf,ext=.pdf,read=.pdf]{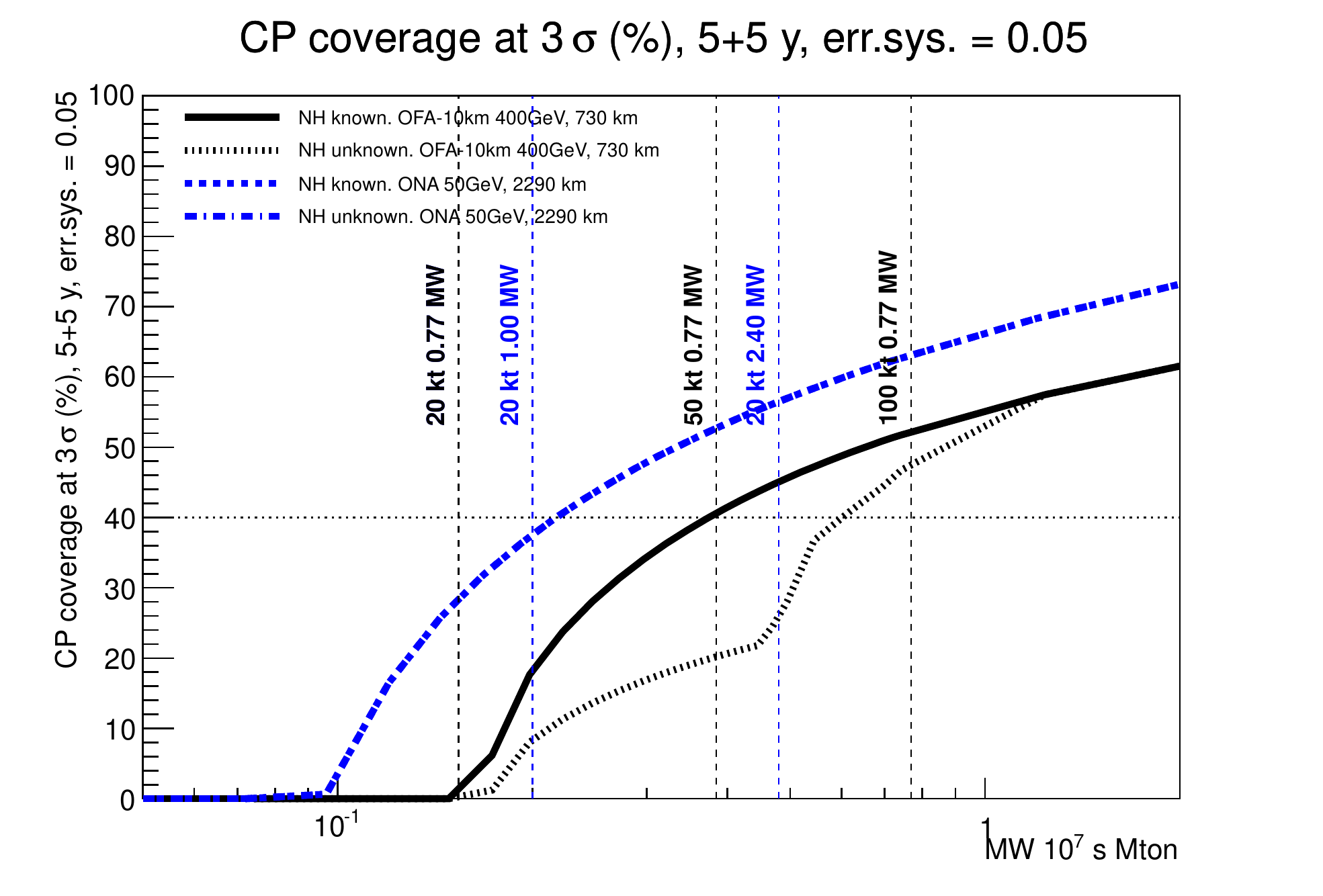} 
\caption{CP coverage for the facilities based on SPS (400 GeV proton
  driver) 10 km off-axis, and on a dedicate driver (50 GeV) at L=2290
  km.}
\label{fig:LNGS-Finland_2}
\end{figure}

Finally, it is interesting to compare the CP reach of medium baselines
with facilities that exploit longer distances to establish the mass
hierarchy through the observation of matter effects. In
Fig.~\ref{fig:compare_baseline} we show the 70\% and 40\% CP coverage
contours as a function of baseline and exposure for three different
beamline configurations. The configurations are labeled $\Phi_{730}$,
$\Phi_{1570}$ and $\Phi_{2300}$: they correspond to beam parameters
optimized (see Sec.~\ref{sec:beamline_detector}) for $L=730, 1570,
2300$~km, respectively. In addition, Fig.~\ref{fig:LNGS-Finland_0}
shows the $3\sigma$ CP coverage for the CERN-to-Phy\"asalmi (2290 km)
facility based on a dedicated 50~GeV proton driver. The same facility
without the dedicated proton driver, i.e. leveraging the existing SPS
complex has a CP coverage described in
Fig.~\ref{fig:LNGS-Finland_2}. As far as mass hierarchy is known in
advance, medium baselines from 700 to 2300 km exhibit similar
performance to establish CP violation in the leptonic sector. In fact,
larger baselines are slightly favoured due to the broader neutrino
energy range and, hence, due to the increase of spectral information.
Spectral information also reduce the dependence of the coverage on the
knowledge of the overall flux normalization and slightly relieve the
deterioration of sensitivity due to
systematics~\cite{Coloma:2012ji,Coloma:2012ut}.
 
\begin{figure}
\centering
\includegraphics[scale=0.4,type=pdf,ext=.pdf,read=.pdf]{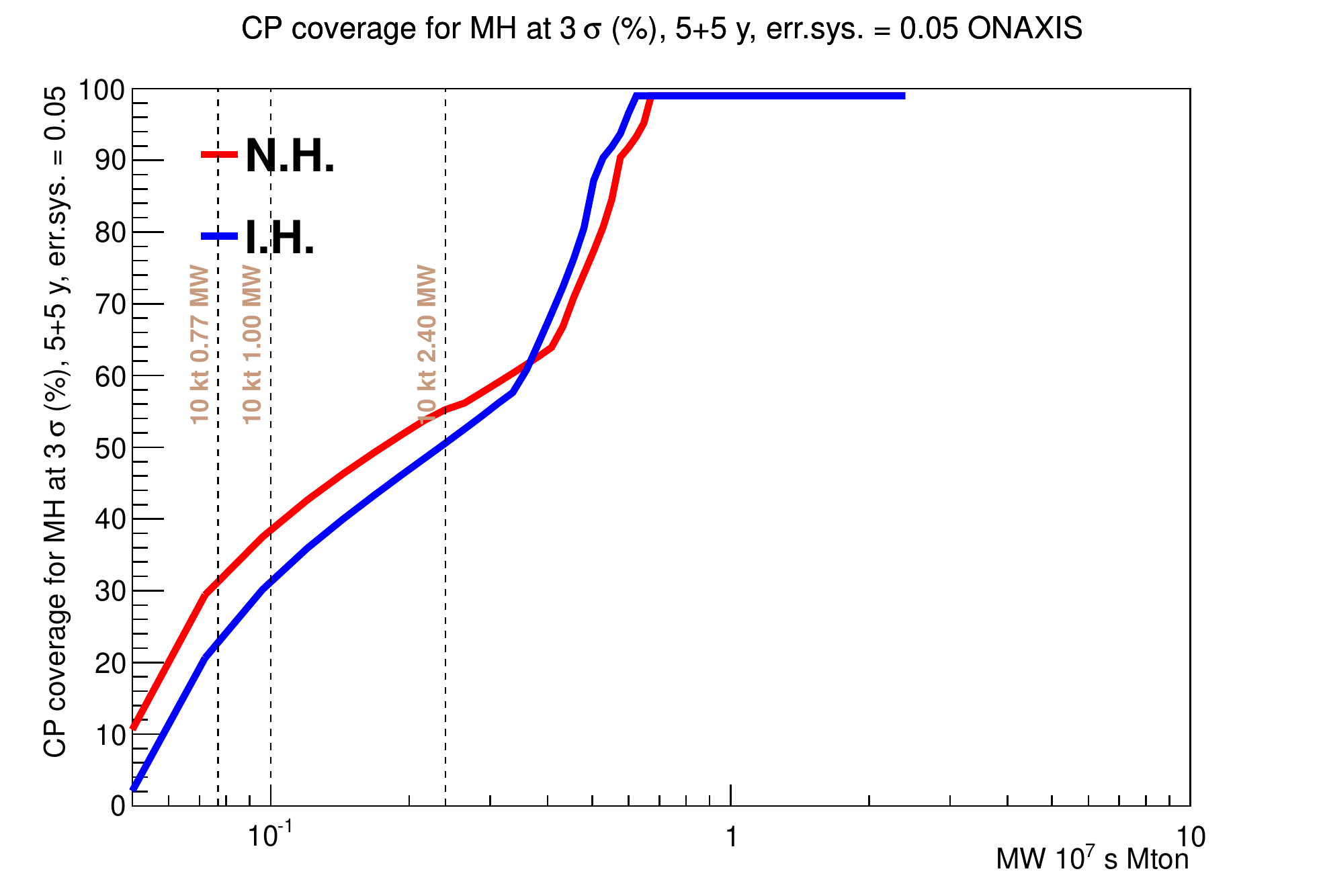}%
\includegraphics[scale=0.4,type=pdf,ext=.pdf,read=.pdf]{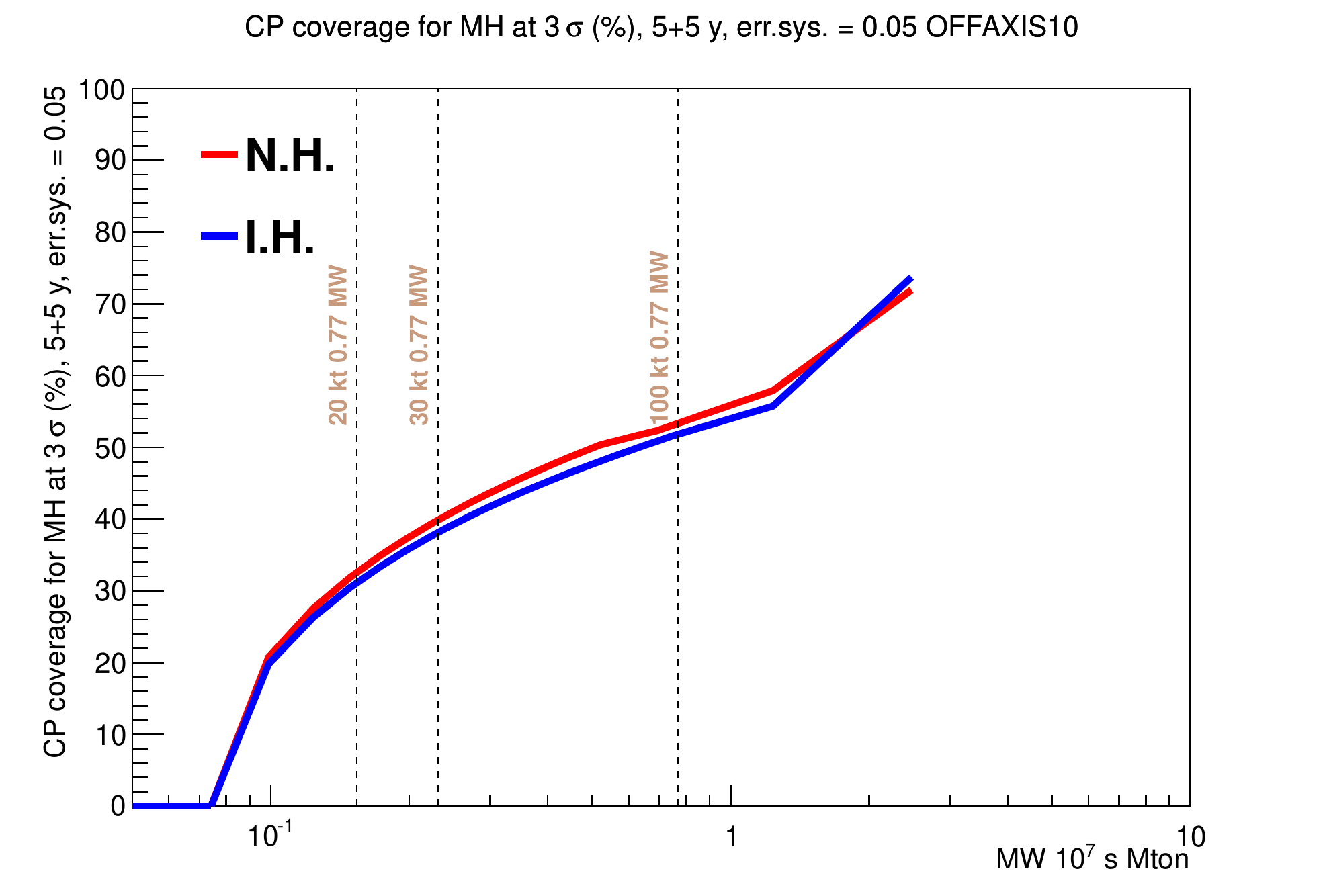}
\caption{Mass hierarchy coverage for the 50~GeV on-axis option (left) and the 400~GeV 10~km off-axis option.}
\label{fig:result6}
\end{figure}

In this
framework, the optimal baseline is somehow in between LNGS and
Phy\"asalmi, i.e. at $L\simeq 1800$~km (Fig.~\ref{fig:compare_baseline}).
At $L\simeq 2000$~km the CP reach is independent of the mass hierarchy
because the sign of $\Delta m^2_{23}$ can be established during data
taking for all values of $\delta$. This is not the case for facilities
at $L\simeq 730$~km, where the mass hierarchy sensitivity hardly
exceeds the one of the NOVA experiment~\cite{NOVA} (see
Fig.~\ref{fig:result6}). Here, the mass hierarchy coverage is defined
as the fraction of possible (true) values of the CP phase $\delta$
where the wrong hierarchy hypothesis is disfavoured at $3\sigma$
level.

\section{Conclusions}

The remarkable size of \tot makes Superbeams the most straightforward
choice for precision physics in the neutrino sector. While the
determination of the mass hierarchy seems at reach for the next
generation of facilities, the study of CP violation remains a major
experimental challenge. Superbeams at medium baselines offer a clean
environment to study CPV, provided that the mass hierarchy is known by
the time of running or the sign ambiguity is resolved employing
spectral information. However, major technological efforts are
required to attain appropriate neutrino fluxes. In this paper we
reconsidered medium baseline superbeams with special emphasis on
CERN-based facilities, which have been re-optimized simulating the
beamline from target to the decay tunnel. Facilities that are based on
the CERN-SPS and exploit the CNGS beamline (off-axis 10~km) reach a
modest CPV coverage in any realistic configuration: at most 52\%
(48\%) with a 100 kton detector in a new shallow laboratory, assuming
the mass hierarchy to be (not) known. Hence, a dedicated high-power
driver is mandatory for CPV studies. Similar considerations hold at
longer baselines. Here we considered specifically the
CERN-to-Phy\"asalmi distance, corresponding to 2290~km, and a new
on-axis neutrino beam from CERN. For a detector mass of 100 kton the
CPV coverage is 62\% and drops to 28\% if the LAr mass is lowered to
20 kton. On the other hand, at large baselines, the size of matter
effects allows for a determination of the mass hierarchy for any value
of $\delta$. Hence, the facility does not suffer from the sign
ambiguity even if the sign of $\Delta m^2_{23}$ is unknown a
priori. The optimal facility to address CPV is an on-axis
configuration (see Figs.~\ref{fig:result4},\ref{fig:sys}) with a
dedicated high power proton driver. A 100 kton detector hosted in a
new underground facility at $L\simeq 730$~km could reach a CPV
coverage of $\sim$ 70\%, fully dominated by systematic uncertainties:
70\% coverage for an overall systematic uncertainty of 5\% and 86\%
coverage if the systematics can be brought down to the (currently
unrealistic) value of 1\%.  Hence, in way similar to what happened
with reactor experiments from Chooz to Daya Bay, novel techniques must
be developed to lower systematics at the percent level in order to
overcome the {\cal O}(70\%) coverage limit.

\section*{Acknowledgments}

We wish to express our gratitude to P. Coloma, E. Fernandez-Martinez, F. Ferroni, A. Masiero,
A. Meregaglia, A. Rubbia, C. Rubbia and L. Votano for useful information and
suggestions.

\end{document}